\documentclass{jfm}

\usepackage{graphicx}
\usepackage{newtxtext}
\usepackage{newtxmath}
\usepackage{natbib}
\usepackage{hyperref}
\hypersetup{
    colorlinks = true,
    urlcolor   = blue,
    citecolor  = blue,
}

\newcommand{\RomanNumeralCaps}[1]
\linenumbers

\title{Restricted Euler dynamics in free-surface turbulence}

\author{Yinghe Qi\aff{1}
  \corresp{\email{yingqi@ethz.ch}},
  Zhenwei Xu\aff{1}
 \and Filippo Coletti\aff{1}}

\affiliation{\aff{1}Department of Mechanical and Process Engineering, ETH Zurich, 8092 Zurich, Switzerland}

\begin{document}

\maketitle

\begin{abstract}
The small-scale velocity gradient is connected to fundamental properties of turbulence at the large scales. By neglecting the viscous and nonlocal pressure Hessian terms, we derive a restricted Euler model for the turbulent flow along an undeformed free surface and discuss the associated stable/unstable manifolds. The model is compared with the data collected by high-resolution imaging on the free surface of a turbulent water tank with negligible surface waves. The joint probability density function (PDF) of the velocity gradient invariants exhibits a distinct pattern from the one in the bulk. The restricted Euler model captures the enhanced probability along the unstable branch of the manifold and the asymmetry of the joint PDF. Significant deviations between the experiments and the prediction are evident, however, in particular concerning the compressibility of the surface flow. These results highlight the enhanced intermittency of the velocity gradient and the influence of the free surface on the energy cascade.

\end{abstract}

\begin{keywords}
Authors should not enter keywords on the manuscript, as these must be chosen by the author during the online submission process and will then be added during the typesetting process (see http://journals.cambridge.org/data/\linebreak[3]relatedlink/jfm-\linebreak[3]keywords.pdf for the full list)
\end{keywords}

\section{Introduction}

The spatial and temporal fluctuations at small scales in fluid turbulence, which are highly non-Gaussian and long-range correlated \citep{mordant2004experimentalintro}, are among the most complex and consequential phenomena in fluid mechanics. The velocity gradient, as well as the velocity increment between two points, is found to be highly intermittent with extreme events occurring more frequently as Reynolds number increases \citep{yeung2015extreme}. One possible mechanism for the enhanced intermittency at small scales is related to the nonlinear self-amplification of the velocity gradient tensor \(A_{ij} = \partial u_{i}/\partial x_{j}\) during its Lagrangian evolution, where \(u_{i}\) is the fluid velocity component in direction \(x_{i}\). This process is directly linked to the classic energy cascade picture in turbulence in which large eddies are assumed to fragment into small eddies \citep{kolmogorov1941local} with energy transferring from large to small scales.

\begin{figure}
    \centering
    \includegraphics[width=\linewidth]{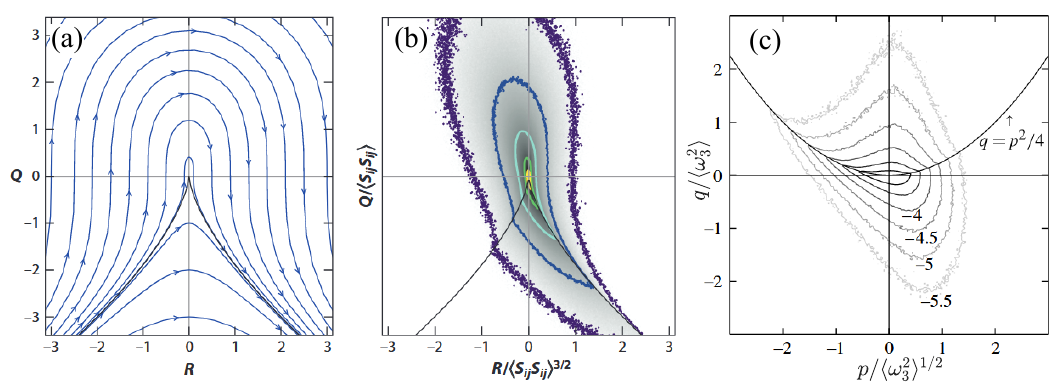}
    \caption{(a) $Q$-$R$ trajectories of the restricted Euler model in 3D turbulence. (b) The joint PDF of $Q$ and $R$ of 3D homogeneous and isotropic turbulence. Here, $S_{ij}$ is the rate-of-strain tensor. Panels (a) and (b) are adapted with permission from \cite{johnson2024multiscale}. (c) The joint PDF of $p$ and $q$ along a 2D section of 3D turbulence. Panels (c) is adapted with permission from \cite{cardesa2013invariants}.}
    \label{fig_intro}
\end{figure}

In addition to the intrinsic strong intermittency, further universal behaviors of \(A_{ij}\) include the negative skewness of the longitudinal velocity gradient \citep{sreenivasan1997phenomenology}, and the preferential alignment between vorticity and the eigenvector associated with the intermediate eigenvalue of the rate-of-strain tensor \citep{kerr1985higher,elsinga2010universal,xu2011pirouette}. Moreover, the joint probability density function (PDF) of the invariants \(Q = - A_{ij}A_{ji}/2\) and \(R = - A_{ij}A_{jk}A_{ki}/3\) is found to exhibit a universal teardrop shape (figure \ref{fig_intro}(b)), with higher probability along the Vieillefosse tail $27R^{2}/4 + Q^{3} = 0$ \citep{vieillefosse1982local,vieillefosse1984internal,meneveau2011lagrangian} for various Reynolds number and flow configurations \citep{soria1994study,chong1998turbulence,luthi2009expanding,elsinga2010evolution,lozano2016multiscale}. These behaviors reflect fundamental processes involved in turbulent flows, e.g., vortex stretching and strain self-amplification.

Dynamic system models from a Lagrangian point of view provide a powerful tool to study the evolution of the velocity gradient and connect dynamics, statistics and structures of turbulence in a unified framework \citep{meneveau2011lagrangian,johnson2024multiscale}. By taking the spatial gradient of the Navier-Stokes equation and neglecting the terms involving the viscous effect and anisotropic pressure Hessian, \citet{vieillefosse1982local,vieillefosse1984internal} and \citet{cantwell1992exact} derived the restricted Euler model:
\begin{equation}\label{eqn_model}
    \frac{dA_{ij}}{dt} + A_{ik}A_{kj} - \left( A_{km}A_{mk} \right)\frac{\delta_{ij}}{3} = 0,
\end{equation}
where \(\delta_{ij}\) is the Kroenecker delta function. Despite the simplifying assumptions, the restricted Euler model captures the self-amplification of the velocity gradient and is remarkably successful at predicting its aforementioned key features, including the teardrop shape observed in the joint PDF of invariants. This model has further inspired subsequent works \citep{girimaji1990diffusion,chevillard2006lagrangian,biferale2007multiscale,chevillard2008modeling,wilczek2014pressure,johnson2016closure} which have attempted to obtain more accurate turbulent statistics by modeling the unclosed terms.

More recently, \citet{cardesa2013invariants} measured the $2\times 2$ reduced velocity gradient tensor \({\widetilde{A}}_{ij}\) along a 2D section of 3D turbulence. The joint PDF of velocity gradient invariants ($p$ and $q$) exhibits a universal teapot pattern for various flow configurations, as illustrated in figure \ref{fig_intro}(c). Although the underlying mechanism leading to the teapot pattern is still not completely understood, the asymmetry of the pattern is found to be associated with the predominance of vortex stretching over compression in 3D turbulence.

In spite of the progress in the understanding and predicting the dynamics of \(A_{ij}\) in 3D turbulence, insights of the reduced velocity gradient tensor \({\widetilde{A}}_{ij}\) in free-surface turbulence, to our best knowledge, is still lacking. Here, we define \({\widetilde{A}}_{ij}\) being the top left \(2 \times 2\) block of the full velocity gradient \(A_{ij}\), with \(i,j = 1,2\) representing the surface-parallel directions. \({\widetilde{A}}_{ij}\) is related to multiple fundamental features of free-surface turbulence, such as surface deformation \citep{babiker2023vortex}, compressible velocity field \citep{boffetta2004large}, exchange of mass between the free surface and the bulk \citep{mckenna2004role,herlina2019simulation}, and its intermittent nature \citep{goldburg2001turbulence,li2004relative}. These features affect a variety of large-scale phenomena including the exchange of gas between the atmosphere and ocean \citep{jahne1998air,veron2015ocean}, the transport of oceanic pollutants such as microplastics \citep{zhang2017transport,mountford2019eulerian,van2020plastic}, and the blooming of phytoplankton \citep{durham2013turbulence,lindemann2017dynamics}. In this work, we developed a restricted Euler model for the reduced velocity gradient tensor \({\widetilde{A}}_{ij}\) at an undeformed free surface, from which the dynamic equations of the invariants are obtained. This model provides new insights on various fundamental features of free-surface turbulence, including its enhanced intermittency observed in laboratory experiments. Deviations from the observations suggest directions to further improve the modeling framework. This paper is organized as follows: in section \ref{sec_model}, we introduce the restricted Euler model for free-surface flows. This model is then compared with experimental data in section \ref{sec_exp}. Section \ref{sec_conclusion} summarizes our findings and draws conclusions.

\section{Restricted Euler model}\label{sec_model}
We begin with adapting the restricted Euler model originally derived for 3D turbulence (equation \ref{eqn_model}). For homogeneous and isotropic free-surface turbulence, the shear-free boundary condition on the free surface requires the derivative of horizontal velocity along the vertical direction to be zero, i.e., $\partial u_1⁄\partial x_3=\partial u_2⁄\partial x_3=0$. We further focus on the situation in which the surface deformation is negligible, which is the case in a wide range of naturally occurring water flows \citep{brocchini2001dynamics}. Then, the no-penetration condition requires the vertical velocity on the free surface to also be zero, i.e., $u_3=0$. This further leads to $\partial u_3/\partial x_1=\partial u_3/\partial x_2=0$. Combining both boundary conditions, we write:
\begin{equation}\label{eqn_bc}
    A_{13} = A_{23} = A_{31} = A_{32} = 0.
\end{equation}
By setting \(i,j = 1,2\) while keeping \(k,m = 1,2,3\) in equation \ref{eqn_model}, the restricted Euler model reduces to the transport equation for \({\widetilde{A}}_{ij}\) which still contains terms such as \(A_{13}\) and \(A_{33}\). By further applying the shear-free and no-penetration boundary conditions on the free surface (equation \ref{eqn_bc}) and the incompressibility condition \(A_{33} = - A_{11} - A_{22} = - \mathrm{tr}\left( {\widetilde{A}}_{ij} \right)\), these terms are eliminated and equation \ref{eqn_model} becomes the restricted Euler equation on the free surface:
\begin{equation}\label{eqn_model_free_surface}
    \frac{d{\widetilde{A}}_{ij}}{dt} + {\widetilde{A}}_{ik}{\widetilde{A}}_{kj} - \left( {\widetilde{A}}_{km}{\widetilde{A}}_{mk} + {\widetilde{A}}_{nn}^{2} \right)\frac{\delta_{ij}}{3} = 0.
\end{equation}
Equation \ref{eqn_model_free_surface} provides a simplified model for the Lagrangian evolution of the reduced velocity gradient tensor in non-wavy free-surface turbulence. Here, \({\widetilde{A}}_{ik}{\widetilde{A}}_{kj}\) is the nonlinear self-amplification term that accounts for the enhanced intermittency, while the term \({\widetilde{A}}_{nn}^{2}\) signals the role of the non-solenoidal nature of the surface flow.

Given the restricted Euler model for \({\widetilde{A}}_{ij}\), the Lagrangian evolution equations for the invariants \(p = - \mathrm{tr}\left( {\widetilde{A}}_{ij} \right) = - {\widetilde{A}}_{11} - {\widetilde{A}}_{22}\) and \(q = \mathrm{det}\left( {\widetilde{A}}_{ij} \right) = {\widetilde{A}}_{11}{\widetilde{A}}_{22} - {\widetilde{A}}_{12}{\widetilde{A}}_{21}\) can also be obtained. The dynamic equation for \(p\) is derived by taking the trace of equation \ref{eqn_model_free_surface}, i.e.,
\begin{equation}\label{eqn_p}
    \frac{dp}{dt} = - \frac{1}{3}p^{2} - \frac{2}{3}q.
\end{equation}
The dynamic equation for \(q\) can be obtained by multiplying equation \ref{eqn_model} by \(A_{ij}\) and then applying the boundary condition on the free surface as well as the incompressibility condition. This leads to the transport equation for the double product term \({\widetilde{A}}_{in}{\widetilde{A}}_{nj}\), i.e.,
\begin{equation}
     \frac{d{\widetilde{A}}_{in}{\widetilde{A}}_{nj}}{dt} + 2{\widetilde{A}}_{in}{\widetilde{A}}_{nk}{\widetilde{A}}_{kj} - \frac{2}{3}\left( {\widetilde{A}}_{km}{\widetilde{A}}_{mk} + {\widetilde{A}}_{nn}^{2} \right)A_{ij} = 0.
\end{equation}
In this equation, the triple product term \({\widetilde{A}}_{in}{\widetilde{A}}_{nk}{\widetilde{A}}_{kj}\) can be rewritten using the Cayley-Hamilton theorem, \({\widetilde{A}}_{in}{\widetilde{A}}_{nk} + p{\widetilde{A}}_{ik} + q\delta_{ik} = 0\), multiplied by \({\widetilde{A}}_{kj}\) to reduce the triple product term to a double one. Then, taking the trace of the transport equation for \({\widetilde{A}}_{in}{\widetilde{A}}_{nj}\) and eliminating the resulting \(dp/dt\) term using equation \ref{eqn_p} leads to the dynamic equation for \(q\):
\begin{equation}\label{eqn_q}
    \frac{dq}{dt} = - \frac{2}{3}p^{3} + \frac{5}{3}pq.
\end{equation}

\begin{figure}
    \centering
    \includegraphics[width=0.5\linewidth]{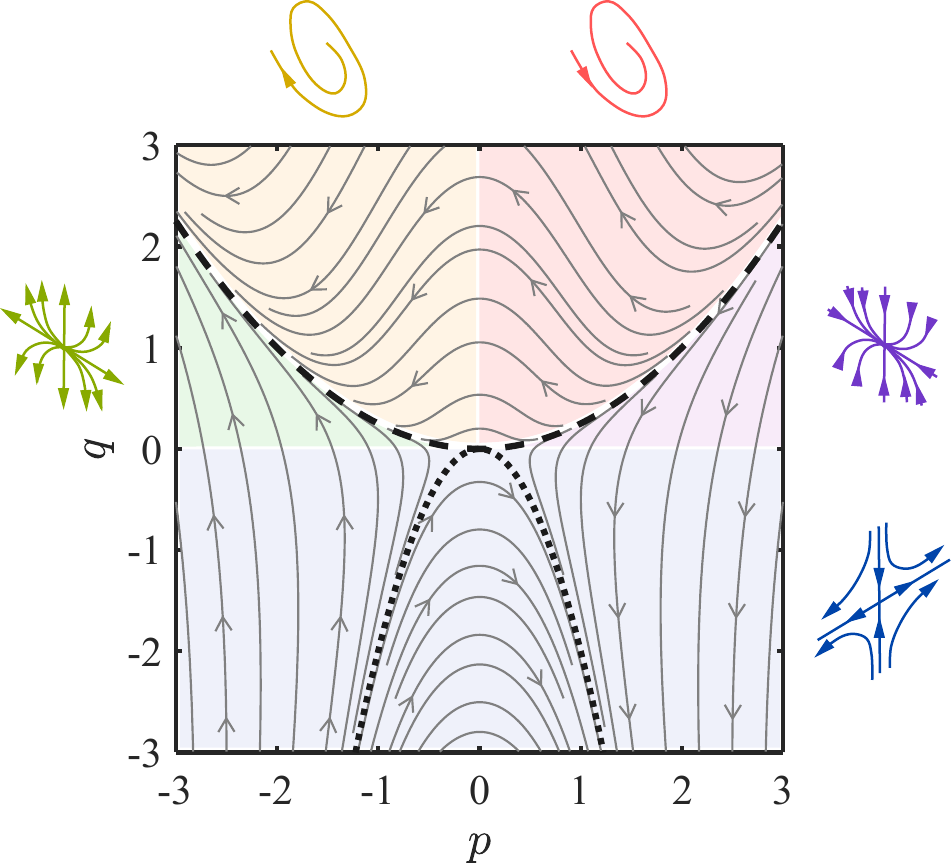}
    \caption{Restricted Euler trajectories (gray lines) calculated based on equations \ref{eqn_p} and \ref{eqn_q}. The arrows mark the direction of the trajectories in the phase portrait. The black dashed line and the black dotted line denote the manifolds $q=p^2/4$ and $q=-2p^2$, respectively. The shades highlight the local flow topology in each region: stable foci (red), unstable foci (yellow), unstable node (green), saddle (blue) and stable node (purple) \citep{perry1987description}.}
    \label{fig_phase_portrait}
\end{figure}

Equations 4 and 5 form a two-dimensional nonlinear dynamic system for the evolution of the invariants. It is evident that there is only one fixed point at the origin \(p = q = 0\). Through this fixed point, two manifolds exist and can be expressed as:
\begin{equation}
    q = p^{2}/4
\end{equation}
and
\begin{equation}
    q = - 2p^{2}.
\end{equation}
The proof of these expressions can be found by comparing the gradient along the manifold, i.e., \(dq/dp\) and \(dq/dp\) calculated by the dynamic system. Evidently, both manifolds exhibit a parabolic form. Moreover, \(q = p^{2}/4\) happens to be the boundary across which the eigenvalues of the reduced velocity gradient tensor \({\widetilde{A}}_{ij}\) change from complex to real and the local flow topology changes between stable/unstable foci and stable/unstable nodes (see figure \ref{fig_phase_portrait}). As suggested by \citet{cantwell1992exact}, this is coincidental: the factor \(1/3\) in equation \ref{eqn_model} is chosen so that the anisotropic pressure Hessian is zero; if a different factor is used, a distinct dynamic system with different manifolds is obtained. We also note that this coincidence is also observed in 3D turbulence, in which the Vieillefosse tail (figure \ref{fig_intro}(a)) also happens to be the boundary that separates the real (corresponding to stable/unstable nodes in 3D) and complex (stable/unstable foci) eigenvalues of the velocity gradient tensor.

It is interesting to examine the stability of these manifolds. For any arbitrary point (\(p_{0}\), \(p_{0}^{2}/4\)) located on the first manifold (\(q = p^{2}/4\)), the dynamics is determined by the Jacobian matrix of the dynamic system:

\begin{equation}\label{eqn_jacob}
    J=
    \begin{bmatrix}
        \dfrac{\partial(dp/dt)}{\partial p} & \dfrac{\partial(dp/dt)}{\partial q}\\[10pt]
        \dfrac{\partial(dq/dt)}{\partial p} & \dfrac{\partial(dq/dt)}{\partial q}
    \end{bmatrix}=
    \begin{bmatrix}
        -\dfrac{2}{3}p_0 & -\dfrac{2}{3} \\[10pt]
        -\dfrac{19}{12}p_0^2 & \dfrac{5}{3}p_0
    \end{bmatrix}.
\end{equation}
We then examine the stability of this point along the normal direction with respect to the manifold, which is determined by calculating the term \({\widehat{e}}_{\bot i}J_{ij}{\widehat{e}}_{\bot j}\), where \({\widehat{e}}_{\bot i}\) is the unit vector normal to the same manifold. The condition \({\widehat{e}}_{\bot i}J_{ij}{\widehat{e}}_{\bot j} > 0\) implies that (\(p_{0}\), \(p_{0}^{2}/4\)) is unstable, and vice versa. By substituting \(q = p^{2}/4\) into \({\widehat{e}}_{\bot i}J_{ij}{\widehat{e}}_{\bot j}\), we can write \({\widehat{e}}_{\bot i}J_{ij}{\widehat{e}}_{\bot j} = {\left( 5p_{0}^{2} + 16 \right)\left( {2p}_{0}^{2} + 8 \right)^{- 1}p}_{0}\). Evidently, the sign of \({\widehat{e}}_{\bot i}J_{ij}{\widehat{e}}_{\bot j}\) is consistent with the sign of \(p_{0}\). This indicates that the first manifold (\(q = p^{2}/4\)) is unstable for \(p > 0\) and stable for \(p < 0\). A similar analysis can be conducted for the other manifold \(q = - 2p^{2}\), and one can write \({\widehat{e}}_{\bot i}J_{ij}{\widehat{e}}_{\bot j} = - \left( 32p_{0}^{2} + 1 \right)\left( 16p_{0}^{2} + 1 \right)^{- 1}p_{0}\). The sign of \({\widehat{e}}_{\bot i}J_{ij}{\widehat{e}}_{\bot j}\) is opposite to the sign of \(p_{0}\), suggesting that this manifold is stable for \(p > 0\) while is unstable for \(p < 0\).

Figure \ref{fig_phase_portrait} shows the \(p\)-\(q\) phase portrait based on the restricted Euler model calculated numerically from equations \ref{eqn_p} and \ref{eqn_q}. The arrows indicate the directions in which the system evolves. Also shown are the two manifold \(q = p^{2}/4\) and \(q = - 2p^{2}\). The trajectories above the parabola \(q = p^{2}/4\) are directed from right to left with a bump around the axis \(p = 0\), and move from left to right below the parabola \(q = - 2p^{2}\). The stability properties of the manifolds are evident by recognizing that the trajectories are directed towards or away from the manifolds depending on the sign of \(p\), consistent with the analysis presented above. The phase portrait also suggests that all initial conditions will eventually proceed to a finite-time singularity along either the left branch of \(q = p^{2}/4\) or the right branch of \(q = - 2p^{2}\). This finite-time singularity along $q=p^2/4$ corresponds to the same singularity along the Vieillefosse tail in 3D turbulence (see figure \ref{fig_intro}(a)) and thus they represent the same physical flow topology with different mathematical manifestations. The presence of finite-time singularities implies that \(p\) and \(q\), as well as \({\widetilde{A}}_{ij}\), will eventually approach infinity via a process of gradient self-amplification, as in 3D turbulence \citep{meneveau2011lagrangian,johnson2024multiscale}. In real flows, viscosity and the pressure Hessian will prevent diverging values of the velocity gradient. However, the presence of two manifolds associated to finite-time singularities suggests that extremely large values of velocity gradients are relatively likely, even more so than in 3D turbulence. Indeed, simulations by \citet{eckhardt2001turbulence} and measurements by \citet{li2004relative} emphasized how free-surface turbulence is characterized by stronger intermittency than 3D turbulence.

The \(p\)-\(q\) phase portrait also predicts how the topology of the local velocity field evolves over time. As illustrated by the schematic and shade in figure \ref{fig_phase_portrait}, it can be seen that stable foci above \(q = p^{2}/4\) in the first quadrant will evolve towards unstable foci. This is consistent with the dynamics in the 3D restricted Euler system (see figure \ref{fig_intro}(a)) in which stable foci also evolve into unstable foci governed by $dR⁄dt=(2/3)Q^2\geq0$. Stable nodes below \(q = p^{2}/4\) in the first quadrant will become saddles, while unstable nodes can only evolve from saddles. Saddles below \(q = - 2p^{2}\) remain such, thus their topology does not significantly change. These behaviors correspond to the intermediate eigenvalue of the three real eigenvalues of the full velocity gradient tensor changing from negative to positive, again equivalent to the dynamics in the 3D restricted Euler system. We emphasize that this insight on the flow topology is based on the restricted Euler model in which the effects of viscosity and anisotropic pressure Hessian are removed.

\begin{figure}
    \centering
    \includegraphics[width=0.9\linewidth]{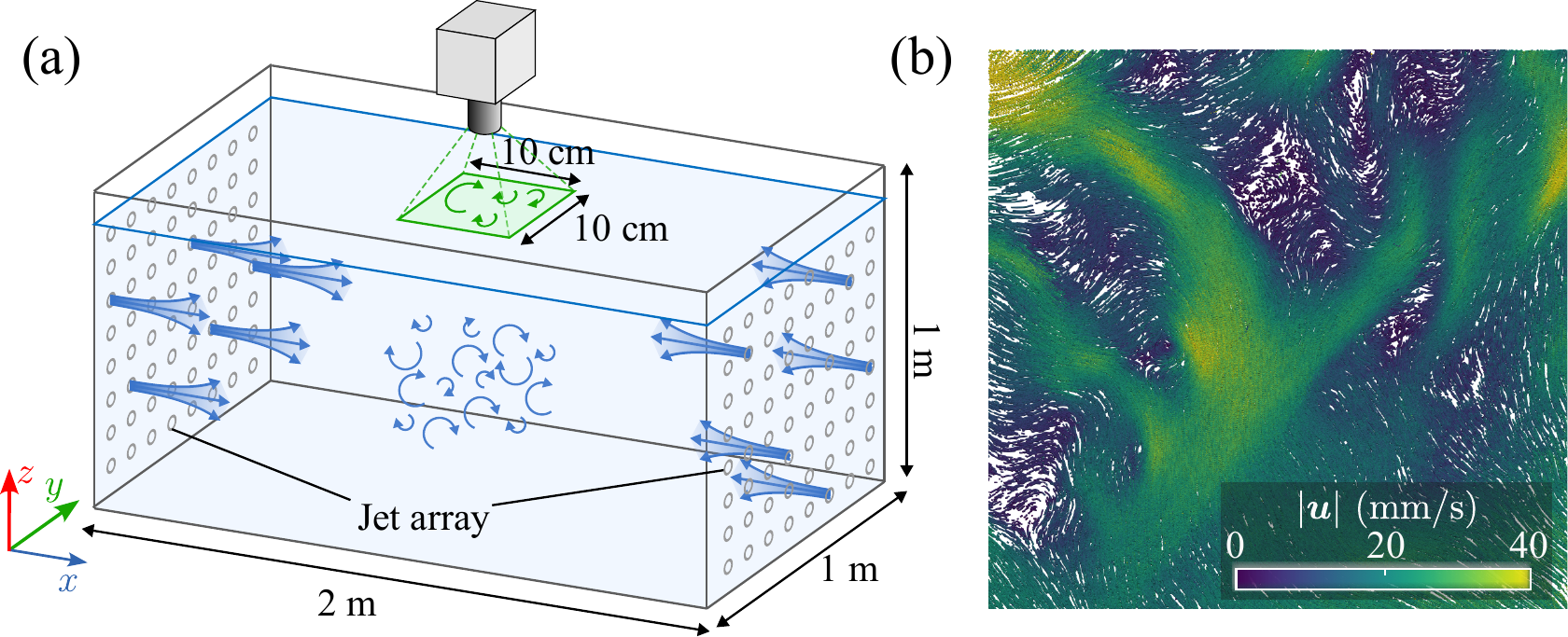}
    \caption{(a) A schematic of the turbulent water tank and camera arrangement. The green shaded area represents the FOV. (b) A snapshot of particle trajectories on the free surface in the FOV at \(Re_{\lambda} = 312\). The trejectories are color coded by the velocity magnitude.}
    \label{fig_tank}
\end{figure}

\section{Experimental setup and results}\label{sec_exp}
To verify the prediction of the restricted Euler dynamics in free surface turbulence, we present results from experiments conducted in a turbulent water tank of dimensions 2\(\times\)1\(\times\)1 m$^3$, illustrated in figure \ref{fig_tank}(a). Two 8×8 arrays of submerged pumps facing each other generate homogeneous turbulence in the center of the tank over a region of about (0.5 m)$^3$. Details regarding this facility can be found in \citet{ruth2024structure} and \citet{li2004relative}. The intensity of the velocity fluctuations \(u_{rms}\) and the dissipation rate of the turbulent kinetic energy \(\epsilon\) are varied in the range \(u_{rms} = 0.02\) to 0.03 m/s and \(\epsilon = 3.82 \times 10^{- 5}\) to \(2.21 \times 10^{- 4}\) m\(^{2}/\)s\(^{3}\) by changing the power supplied to each pump. This leads to a range of Taylor Reynolds number \(Re_{\lambda} = 207\) to \(312\). We note that strong surface deformation due to the sub-surface turbulence could bring difficulties and large uncertainty in the surface velocity measurement. However,in this regime, the surface remains essentially flat, with only sub-millimeter deformations due to the sub-surface turbulence. Therefore, such surface deformation is neglected and will not significantly affect the result.

The free surface of the water is maintained at 8 cm above the axis of the jets at the top row of the arrays. As depicted in figure \ref{fig_tank}(a), the surface motion is captured by a downward looking CMOS camera placed about 0.31 m above the surface. This is operated at 400 fps, with a resolution of 1664\(\times\)1600 pixels over a 10\(\times\)10 cm\textsuperscript{2} field of view (FOV) illuminated by two LED panels. To resolve the small-scale dynamics, the surface motion is characterized by tracking floating hollow glass microspheres (63--75 µm in diameter and 0.31g/cm$^3$ in density), at a high concentration about 120 particles/cm$^2$. Individual particles are identified and tracked via an in-house particle tracking velocimetry (PTV) code based on the nearest-neighbour algorithm \citep{petersen2019experimental}. The velocity is obtained by convolving the trajectories with the first derivative of a temporal Gaussian kernel, whose width is comparable to the smallest time scales of the flow \citep{mordant2004experimental}. Figure \ref{fig_tank}(b) shows an example of trajectories obtained over a series of 25 successive images.

The reduced velocity gradient tensor \({\widetilde{A}}_{ij}\) at each particle position on the free surface is obtained by performing a least-square fit based on the velocity of particles within a search radius \(R_{s}\) \citep{pumir2013tetrahedron,qi2022fragmentation}. When \(R_{s}\) is large, the calculated velocity gradient will be coarse-grained; while small \(R_{s}\) leads to a limited number of surrounding particles, thus larger uncertainty. The value is thus determined following an approach similar to the one for the PTV kernel \citep{mordant2004experimental}: \(R_{s}\) is chosen as the smallest value above which the standard deviation of \({\widetilde{A}}_{ij}\) decays exponentially with it. Following this approach, \(R_{s} = 2.5\) mm is used which yields on average 40 particles within the search radius. This $R_s$ is comparable with the Kolmogorov length scale defined on the free surface and the results are insensitive to the precise value of the search radius. More details can be found in \citet{qi2024small}.
\begin{figure}
    \centering
    \includegraphics[width=\linewidth]{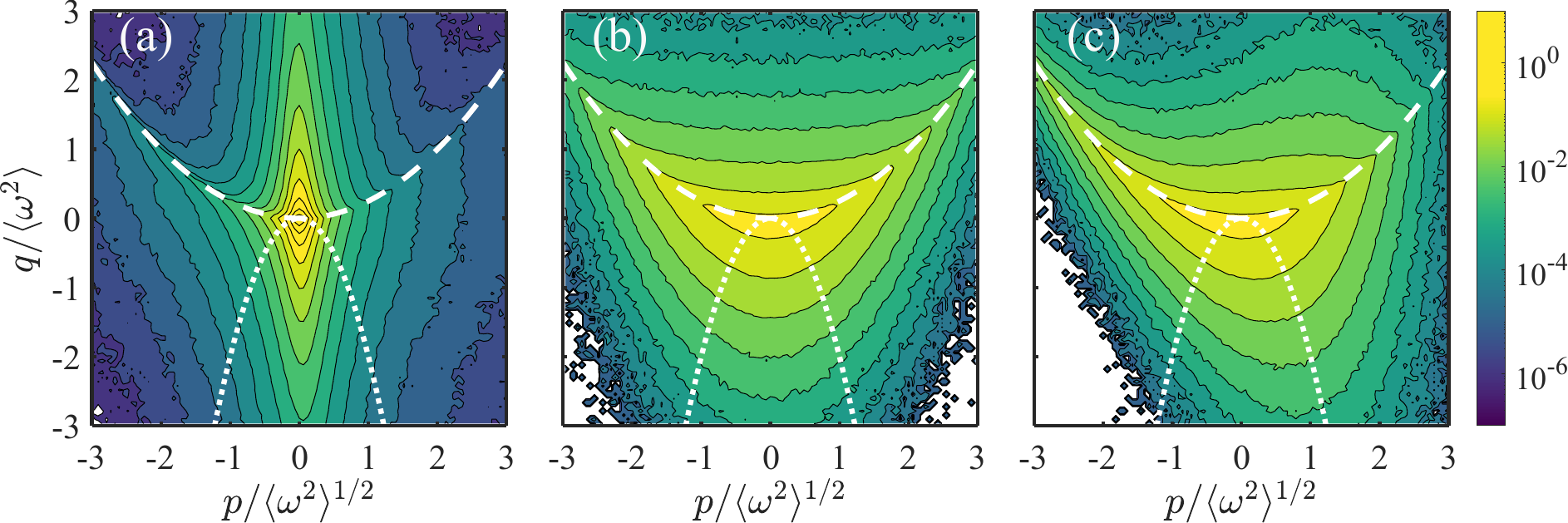}
    \caption{(a) The joint PDF of normalized \(p\) and \(q\) at \(Re_{\lambda} = 312\) based on the experimental data. The logarithmic contours range from \(10^{- 7}\) to \(10\) with adjacent contours being separated by half a decade. (b) The initial joint PDF of normalized \(p\) and \(q\) for the Monte-Carlo simulation. (c) The joint PDF of normalized \(p\) and \(q\) obtained from the Monte-Carlo simulation at $\langle\omega^2\rangle^{1/2} t=0.15$. For (b--c), the logarithmic contours range from \(10^{- 6}\) to \(1\) with adjacent contours being separated by half a decade. In all the three panels, the white dashed lines mark \(q = p^{2}/4\), and white dotted lines mark \(q = - 2p^{2}\).}
    \label{fig_jpdf}
\end{figure}

With \({\widetilde{A}}_{ij}\) obtained from the experiment, the invariants \(p\) and \(q\) are calculated and their joint PDFs are shown in figure \ref{fig_jpdf}(a) for \(Re_{\lambda} = 312\). Both invariants are normalized by the vorticity of the surface flow, \(\omega = {\widetilde{A}}_{21} - {\widetilde{A}}_{12}\). The dashed line and dotted line mark the manifolds \(q = p^{2}/4\) and \(q = - 2p^{2}\) obtained from the reduced Euler representation. We note the distribution of the joint PDF does not change significantly among various cases. Thus, the following discussion should apply for at least the Reynolds number range considered in this work.

Remarkably, the contours in the \(p\)-\(q\) joint PDF show similarities with the trajectories in the analytical phase portrait shown in figure \ref{fig_phase_portrait}. In particular, a relatively high probability is found along the manifold \(q = p^{2}/4\), especially its left branch, highlighting the role of strain self-amplification which results in exceptionally strong intermittency. Quantitatively, the kurtosis of the velocity gradients along the surface ($\widetilde{A}_{11}$) is approximately 9.5, which is higher than the one for 3D homogeneous turbulence at similar $Re_\lambda$ \citep{gylfason2004intermittency}. The asymmetry of the distribution reflects the stability properties of this manifold, i.e., the left branch is stable as discussed above and thus exhibits higher probability. Although the manifold \(q = - 2p^{2}\) does not leave an obvious footprint, the relatively higher probability in the fourth quadrant than the one in the third quadrant is consistent with the stability of its right branch. The fact that no clear tail is observed along this manifold still requires further investigation.

To better compare the joint PDF with the restricted Euler model, a Monte-Carlo simulation is performed. A large ensemble of $\mathcal{O}(10^7)$ initial sets of invariants is considered and each of these sets is evolved following equations \ref{eqn_p} and \ref{eqn_q}. The initial values of invariants are obtained from the reduced velocity gradient tensor, each element of which follows a Gaussian distribution with the same variance and zero mean, as illustrated in figure \ref{fig_jpdf}(b). We note that as the restricted Euler system is divergent, the predicted joint PDF does not converge to a statistically stationary state. Thus, in figure \ref{fig_jpdf}(c), the predicted joint PDF at an intermediate time instant ($\langle\omega^2\rangle^{1/2} t=0.15$, where $t$ indicates time) is used to compare with the experimental data, though the features remain qualitatively similar at different times.

In figure \ref{fig_jpdf}, large deviations between the experimental and the predicted joint PDF are evident, in particular the high probability around $p=0$ in the former. This is associated to low levels of the compressibility ratio $\mathcal{C}=\langle(\widetilde{A}_{ii})^2\rangle/\langle(\widetilde{A}_{ij})^2\rangle=0.013$--$0.024$ found in the measurements. This is much lower than $\mathcal{C}=0.45$--$0.5$ observed by \citet{cressman2004eulerian} who forced turbulence at much deeper distance under the free surface with $Re_\lambda=100$--$140$. It is possible that differences in the details of the forcing scheme impact the state of the surface flow. We note, however, that the compressibility ratio we measure is largely unaffected by changing the depth of the forcing. Despite the large deviations, the predicted joint PDF still successfully reproduces some features of the experimental data; in particular, the higher probability along the left branch of $q=p^2/4$ and the slightly higher probability in the fourth quadrant than the one in the third quadrant. The manifold $q=-2p^2$ does not show clear footprint as in the experimental data, indicating that this manifold might not affect the dynamics of the reduced velocity gradient as much as the manifold $q=p^2/4$.

Given the asymmetry observed in the measured joint PDF (figure \ref{fig_jpdf}(a)), it is informative to examine the mean of both invariants. As the gradients of the mean velocity components are weak \citep{li2004relative}, the average of \(p\) is expected to be approximately zero \(\langle p\rangle = - \partial\langle u_1\rangle/\partial x_1 - \partial\langle u_2\rangle/\partial x_2 = 0\). In addition, given the homogeneity of the free-surface turbulence, the cross product of velocity gradient also satisfies \(\langle(\partial u_1/\partial x_1)(\partial u_2/\partial x_2)\rangle = \langle(\partial u_1/\partial x_2)(\partial u_2/\partial x_1)\rangle\) \citep{george1991locally}. Substituting this relation into the definition of \(q\) results in \(\langle q\rangle = \langle(\partial u_1/\partial x_1)(\partial u_2/\partial x_2)\rangle - \langle(\partial u_1/\partial x_2)(\partial u_2/\partial x_1)\rangle = 0\). Therefore, both invariants are expected to have zero mean, \(\langle p\rangle = \langle q\rangle = 0\). This is approximately confirmed by the present data, which gives \(\left\langle p \right\rangle/\left\langle \omega^{2} \right\rangle^{1/2} = 1.7 \times 10^{- 3}\) and \(\left\langle q \right\rangle/\left\langle \omega^{2} \right\rangle = - 0.035\).

It is also noted that the joint PDF of invariants (figure \ref{fig_jpdf}(a)) displays a distinct pattern from the one obtained from generic 2D sections of 3D turbulence \citep{cardesa2013invariants}. The teapot shape (as in figure \ref{fig_intro}(c)) shows a much more pronounced asymmetry compared to the present free-surface case. This asymmetry, quantified by the inequality \(\left\langle pq \right\rangle < 0\), is found to be connected to the predominance of vortex stretching over vortex compression in 3D turbulence. In particular, they found that \(\left\langle pq \right\rangle/\left\langle \omega^2 \right\rangle^{3/2}\) ranged between $-0.044$ and $-0.067$ for 3D turbulence, while in the present case we measure \(\left\langle pq \right\rangle/\left\langle \omega^2 \right\rangle^{3/2} = - 1.8 \times 10^{- 3}\). Therefore, we deduce that the weaker asymmetry observed here is due to the no-shear-stress boundary condition which eliminates the vortex stretching along the free-surface directions. This is crucial for the dynamics, as vortex stretching is a major factor in the inter-scale energy transfer in 3D turbulence \citep{johnson2020energy,johnson2021role,davidson2015turbulence}. This may have far-reaching consequences for the energy cascade associated to the dynamics along the free surface, which has been found to exhibit inverse inter-scale fluxes from scales small to large \citep{pan1995numerical,lovecchio2015upscale}. Moreover, surface-attached vortices are linked to surface-divergence events \citep{babiker2023vortex} which are crucial for gas transfer to and from the liquid \citep{jahne1998air,herlina2019simulation}. Inspired by these considerations, further studies are warranted to conduct a complete description of the surface flow topology and dynamics.

\section{Conclusion}\label{sec_conclusion}
In this work, we have developed a restricted Euler model for the reduced velocity gradient tensor in free-surface turbulence by simplifying the 3D restricted Euler model using free-surface boundary conditions. Two manifolds associated to finite-time singularities appear in the phase portrait that describes the dynamics of the invariants \(p\) and \(q\), highlighting the intrinsic intermittent nature of \({\widetilde{A}}_{ij}\). The model is compared with experimental data obtained in a turbulent water tank with quasi-flat free surface, in which the surface velocity gradient is obtained by tracking highly concentrated floating microspheres. The \(p\)-\(q\) joint PDF shows a distinct pattern which differs significantly from the one measured in 2D sections of 3D turbulence. Some features of the joint PDF, including the high probability along the unstable branch of $q=p^2/4$ and the asymmetry of the distribution, are predicted by the restricted Euler model. This study provides experimental evidence as well as a theoretical basis for the enhanced intermittency in free-surface turbulence.

In spite of the success of the current model in predicting certain features, large deviations between the experimental and predicted joint PDFs are evident, particularly the high probability around $p=0$, associated to a weak surface compressibility observed in the experiments which deserves further investigation. In addition, the magnitude of the skewness is greatly overpredicted. We remark that the current model is inherently not convergent (i.e., does not reach a steady state) due to the lack of pressure Hessian and viscous effect. Further improvements of the model to take into account these terms are needed to guarantee convergent and more accurate predictions, and to advance the understanding on the evolution of the reduced velocity gradient tensor on the free surface.

\backsection[Funding]{Funding from the Swiss National Science Foundation (project \# 200021-207318) is gratefully acknowledged.}

\backsection[Declaration of interests]{The authors report no conflict of interest.}

\backsection[Data availability statement]{All the data supporting this work are available from the corresponding author upon reasonable request.}

\backsection[Author ORCIDs]{Y. Qi, https://orcid.org/0009-0004-9858-9411; Z. Xu, https://orcid.org/0009-0006-7044-3992; F. Coletti, https://orcid.org/0000-0001-5344-2476}

% \backsection[Author contributions]{Authors may include details of the contributions made by each author to the manuscript'}

\bibliographystyle{jfm}
\bibliography{ref}

\end{document}